\documentclass[aps,prl,twocolumn,nofootinbib,superscriptaddress,floatfix]{revtex4-1}

\usepackage{amsmath}
\usepackage{bbold}
\usepackage{amsfonts}
\usepackage{graphicx}
\usepackage{verbatim}
\usepackage{tikz}
\usepackage{color}
\usepackage{layouts}
\usepackage{bbm}
\usepackage[]{lineno}

\begin{document}
\title{Controllability in tunable chains of coupled harmonic oscillators}

\author{L.~F.~Buchmann}
\affiliation{Department of Physics and Astronomy, Aarhus University, Ny Munkegade 120, DK 8000 Aarhus C, Denmark}
\affiliation{Institute of Electronic Structure and Laser, FORTH, GR-71110 
Heraklion, Crete, Greece}
\author{K.~M\o lmer}
\affiliation{Department of Physics and Astronomy, Aarhus University, Ny Munkegade 120, DK 8000 Aarhus C, Denmark}
\author{D.~Petrosyan}
\affiliation{Institute of Electronic Structure and Laser, FORTH, GR-71110 
Heraklion, Crete, Greece}

\date{\today}
\begin{abstract}
We prove that temporal control of the strengths of springs connecting $N$ harmonic oscillators in a chain provides complete access to all Gaussian states of $N-1$ collective modes. The proof relies on the construction of a suitable basis of cradle modes for the system. An iterative algorithm to reach any desired Gaussian state requires at most $3N(N-1)/2$ operations. We illustrate this capability by engineering squeezed pseudo-phonon states -- highly non-local, strongly correlated states that may result from various nonlinear processes. Tunable chains of coupled harmonic oscillators can be implemented by a number of current state-of-the-art experimental platforms, including cold atoms in lattice potentials, arrays of mechanical micro-oscillators, and coupled optical waveguides. 
\end{abstract}

\maketitle


Chains of coupled harmonic oscillators are simple, yet non-trivial systems that can be used to study complex physical phenomena. Their use goes back to at least Schr\"odinger, who derived thermodynamic properties of solids in 1914~\cite{schrodinger1914}, and continues to this day, with investigations including statistical and mathematical physics~\cite{dudnikova2017, jeugt2008}, transport properties in nanowires~\cite{mingo2003} and non-classical effects in open quantum systems~\cite{dissipative}. 
Many novel experimental platforms implement harmonic oscillator chains in the quantum regime, including ion crystals~\cite{ionchain, ionchain2}, nanomechanical arrays~\cite{omarray1}, coupled waveguides~\cite{waveguide1} and ultracold atoms trapped in optical lattices~\cite{omarray2}. These implementations can cover a wide range of parameters to simulate and study non-equilibrium phenomena~\cite{thermodynchains} and to employ harmonic oscillator chains for various quantum information applications. 
Extensive research of, e.g., the dynamics of entanglement~\cite{plenio2002, reznik2004, brukner2006, plenio2004, tonni2017, vitali2015, morigi2014} and quantum state transfer in chains~\cite{giorgos, plenio2006, plenio2005, plenio2005_2} have led to several quantum information applications of Gaussian states~\cite{gaussqis, cluster}, including universal quantum computation~\cite{universalCVqcomp}, quantum secret sharing~\cite{freqcombs}, cloning and teleportation~\cite{cloning}. 

It is often assumed that the couplings between neighboring oscillators in the chain are uniform or follow a certain pattern suitable for the desired application. Yet, in many realizations of oscillator chains, such as atoms in arrays of microtraps~\cite{saffmann2010, browaeys2016,pohl2014,buchmann2017}, sequentially coupled optomechanical cavities~\cite{omarray1} or ions in Coulomb crystals~\cite{ionchain}, one can tune the strength and time-dependence of individual couplings.

Whether or not a tunable chain of harmonic oscillators is suitable for universal quantum computation with continuous variables~\cite{universalCVqcomp}, is capable of quantum teleportation~\cite{cloning} or can serve as a sensor of spatially extended weak fields or surfaces~\cite{martin2012, spreeuw2017} depends critically upon the controllability of the system, which determines the set of states that can be prepared by manipulating individual springs. Consequently, the controllability of harmonic oscillator chains has received careful attention from the theoretical community. The circumstances under which the rank criterion may be used on such systems has been established~\cite{genoni1} and the set of reachable states under parametric interactions has been characterized~\cite{genoni2}. 

Here we consider a chain of harmonic oscillators and assume that each oscillator consists of the same physical system and each spring coupling neighboring oscillators is realized by the same underlying mechanism, e.g. switchable interatomic forces or fiber coupling between optomechanical cavities. Hence, if one of the couplings can be tuned in strength and time, so can all the others. In this sense, our conditions are minimal, because any less control would imply an oscillator that is not coupled to the remaining chain.
We prove that the control of the time-dependent springs between any two neighboring sites of a chain of $N$ oscillators gives complete access to all pure Gaussian states of $N-1$ modes of the system. We present an explicit algorithm for constructing any desired state and demonstrate it with the engineering of an $N$ body squeezed state of a ``pseudo-phonon'' mode. To streamline the analytic treatment, we introduce cradle modes of the finite system that permit an inductive proof and aids in the construction of a recursive scheme to reach a given target state.

Consider a chain of $N$ degenerate harmonic oscillators with frequency $\omega$.
Any two neighboring oscillators $n$ and $n+1$ are connected by a spring with tunable strength 
$\Omega_n(t)$. The Hamiltonian reads ($\hbar=1$)
\begin{linenomath*}
\begin{align}\label{hamiltonian}
H=\omega\sum_{n=1}^N\left(\frac{\hat{p}_n^2}{2}+\frac{\hat{x}_n^2}{2}\right)+\sum_{n=1}^{N-1}\Omega_n(t)\left(\hat{x}_{n}-\hat{x}_{n+1}\right)^2,
\end{align}
\end{linenomath*}
where $\hat{x}_n$ and $\hat{p}_n$ are the dimensionless position and momentum operators for the $n$th oscillator, satisfying $[\hat{x}_n,\hat{p}_m]=i\delta_{nm}$. We neglect the effects of dissipation, assuming that relaxations occur on time-scales much longer than the maximum of $\Omega_n^{-1}$.
Since the Hamiltonian is quadratic in position and momentum operators, their first and second order moments decouple. We are interested in the mechanical quantum fluctuations of the system and disregard the first moments. 
The dynamics governed by Hamiltonian (\ref{hamiltonian}) preserves the Gaussian character of states and the system is completely characterized by a vector of Heisenberg operators $\hat{\mathbf{q}}=(\hat{x}_1,\hat{p}_1,\ldots,\hat{x}_N,\hat{p}_N)^\top$. The evolution is described by matrices acting on $\hat{\mathbf{q}}$, the only restriction being the conservation of commutation relations. These matrices form the symplectic group over the reals $\mathrm{Sp}(2N,\mathbbm{R})$. Below we prove that by appropriate manipulation of the time dependence of couplings $\Omega_n$ we can access all matrices in $\mathrm{Sp}(2(N-1),\mathbbm{R})$.

To simplify the proof, we assume that no two neighboring springs are turned on at the same time. When a particular spring $\Omega_n$ is turned on, it couples the motion of oscillators $n$ and $n+1$. If, in addition, the strength of the spring is modulated at twice the bare oscillator frequency, 
$\Omega_n = \bar{\Omega}_n (t) [1+A(t) \cos(2 \omega t+\phi)] $, the interaction is parametric~\cite{buchmann2017}. The two oscillators with the relative position and momentum coordinates $\hat{\mathbf{r}}_n^-= \frac{1}{\sqrt{2}}(\hat{x}_{n}-\hat{x}_{n+1},\hat{p}_{n}-\hat{p}_{n+1})^\top$ are then driven into an entangled state. In the phase space spanned by these relative coordinates, parametric coupling results in a squeezed ellipse for the quasi-probability distribution of the state. The orientation of this ellipse, i.e. the squeezing angle, is determined by the phase $\phi$ of modulation. The squeezing magnitude and angle completely characterize the state of the oscillator corresponding to $\hat{\mathbf{r}}_n^-$. Thus tuning $\Omega_n(t)$ gives access to all Gaussian states of the relative motion of oscillators $n$ and $n+1$. Formally, this is described by the ability to create an arbitrary symplectic matrix $S$ acting on the relative coordinates,
\begin{equation}
\hat{\mathbf{r}}_n^-\to \exp\left(\alpha s_1+\beta s_2 +\gamma s_3\right)\hat{\mathbf{r}}_n^-\equiv S \hat{\mathbf{r}}_n^-,
\end{equation}
where $\alpha,\beta,\gamma$ are real numbers and 
\begin{linenomath*}
\begin{align}
  s_1=\left(\begin{array}{cc}
              1&0\\
              0&-1
        \end{array}\right),
  s_2=\left(\begin{array}{cc}
              0&1\\
              1&0
            \end{array}\right),
  s_3=\left(\begin{array}{cc}
              0&-1\\
              1&0
        \end{array}\right),
\end{align}
\end{linenomath*}
are the generators of the symplectic group of order 2. 
In contrast, the phase space distribution of the sum coordinate $\hat{\mathbf{r}}_n^+= \frac{1}{\sqrt{2}}(\hat{x}_{n}+\hat{x}_{n+1},\hat{p}_{n}+\hat{p}_{n+1})^\top$ are not changed by $\Omega_n$. More generally, for any combination of couplings $\Omega_n$, the Hamiltonian (\ref{hamiltonian}) commutes with the total displacement $\hat{x}_\text{s}=\frac{1}{\sqrt{N}}\sum_n^N\hat{x}_n$ and its conjugate momentum, which limits the controllability in the system~\cite{buchmann2017}. 

\begin{figure}
\includegraphics[width=0.6\columnwidth]{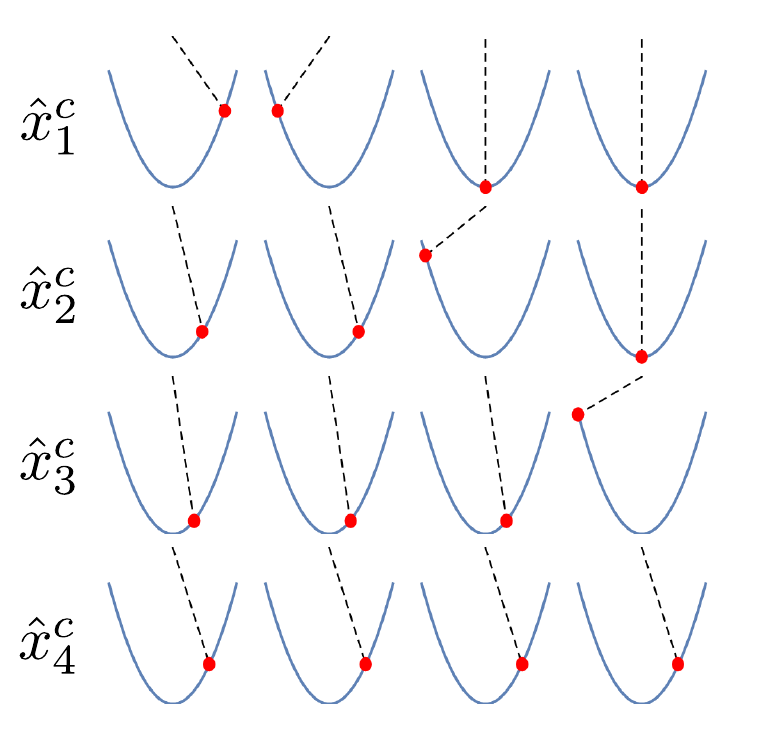}
\caption{Sketches of the cradle modes and total discplacement of the system for $N=4$.}
\label{fig:cradlemodes}
\end{figure}

To describe the total system, we introduce a basis that keeps the matrices for coupled pairs of oscillators simple, conserves the canonical commutation relations between the distinct modes, and separates the invariant $\hat{x}_s$. A physically intuitive basis is given by the ``cradle'' coordinates 
\begin{linenomath*}
\begin{align}
\hat{x}^c_j=\sqrt{\frac{j}{j+1}}\left(\frac{1}{j}\sum_{n=1}^{j}\hat{x}_n-\hat{x}_{n+1}\right),
\end{align}
\end{linenomath*}
and analogously for $\hat{p}^c_j$, with $j$ taking values $1,\ldots, N-1$. Each cradle mode $j$ describes the motion involving only the first $j+1$ oscillators, with the first $j$ oscillators moving in phase with the same amplitude $\frac{1}{\sqrt{j(j+1)}}$, and oscillator $j+1$ moving out of phase with amplitude $\sqrt{\frac{j}{j+1}}$, as illustrated in Fig.~\ref{fig:cradlemodes}. The cradle modes are orthogonal and obey the commutation relations $[\hat{x}^c_i,\hat{p}^c_j]=i\delta_{ij}$. Since the last mode $\hat{x}_N^c\equiv\hat{x}_s$, corresponding to the total displacement, is not affected by the interactions, we may exclude it from our analysis. We can then characterize the entire chain by the vector of $\tilde{N}\equiv N-1$ pairs of Heisenberg operators $\hat{\mathbf{q}}_c=(\hat{x}^c_1, \hat{p}^c_1, \ldots \hat{x}_{\tilde{N}}^c, \hat{p}^c_{\tilde{N}})^\top$. 

Acting on oscillators $n$ and $n+1$ with matrix $S$, $\hat{\mathbf{r}}^-_n\to S \hat{\mathbf{r}}^-_n$, changes the state of the system to $\hat{\mathbf{q}}_c\to D_n(S) \hat{\mathbf{q}}_c$, where the symplectic $2\tilde{N}\times 2\tilde{N}$ matrices $D_n(S)$ are found from the transformations between the cradle and single-oscillator bases. The matrix $D_{n=1}(S)$ is given by $S$ in the upper left $2\times 2$ corner, unity on the diagonal and zeroes everywhere else. For $n>1$, $D_n(S)$ remain block-diagonal
\begin{linenomath*}
\begin{align}
D_n(S)=\left(\begin{array}{c|c|c}
\mathbb{1}_{2n-4}&0&0\\
\hline
0&P_n(S)&0\\
\hline
0&0&\mathbb{1}_{2\tilde{N}-2n}\\
\end{array}\right),
\end{align}
\end{linenomath*}
with $\mathbb{1}_j$ the identity matrix of dimension $j$ and 
\begin{linenomath*}
\begin{align}\label{Pk}
P_n(S)=\left(\begin{array}{cc}
\frac{n+1}{2n}\mathbb{1}_2+\frac{n-1}{2n}S&\frac{\sqrt{n^2-1}}{2n}(\mathbb{1}_2-S)\\
\frac{\sqrt{n^2-1}}{2n}(\mathbb{1}_2-S)&\frac{n-1}{2n}\mathbb{1}_2+\frac{n+1}{2n}S
\end{array}\right).
\end{align}
\end{linenomath*}
The Lie algebra spanned by the matrix logarithms $d_n(s)\equiv\log[D_n(e^{s})]$ determines the states that can be created by sequential applications of symplectic operations on the relative coordinates of neighboring oscillators. The simple form of the transformation matrices makes it easy to explicitly calculate the matrix logarithms. In particular, the generator $d_1(s)$ corresponding to $D_1(S)$ is simply the matrix logarithm of $S$ in the upper left corner and zeroes everywhere else. For other modes $n>1$, the only non-vanishing entries in the generator $d_n(s)$ is a $4\times 4$ block on the diagonal 
\begin{linenomath*}
\begin{align}\label{pksc}
 p_n(s)=\frac{1}{2n}\left(\begin{array}{cc}
                             (n-1)s & -\sqrt{n^2-1}s\\
                             -\sqrt{n^2-1}s&(n+1)s
                           \end{array}\right).
\end{align}
\end{linenomath*}

For $n=2$, the generators $d_1(s_i)$ and $d_2(s_i)$ provide 6 linearly independent generators. Their commutator $[d_1(s_i),d_2(s_j)]$, which can be obtained from
\begin{linenomath*}
\begin{align}\label{commutator}
  \left[\left(\begin{array}{cc}s_i & 0\\0&0\end{array}\right),p_n(s_j)\right]=
                                           \left(\begin{array}{cc}
                                                   \frac{(n+1)}{2n}[s_i,s_j]&-\frac{\sqrt{n^2-1}}{2n}s_is_j\\
                                                   \frac{\sqrt{n^2-1}}{2n}s_js_i           &0
                                                 \end{array}\right),
\end{align}
\end{linenomath*}
yields one more linearly independent matrix for $i=j$ and three additional ones for $i\neq j$. We thus obtain 10 linearly independent generators for the first two cradle modes. Since the dimension of the Lie-algebra for the symplectic group of order $2n$ is $\dim(\mathfrak{sp}(2n,\mathbbm{R}))=n(2n+1)$, all symplectic transformations involving the first two cradle modes can be accessed with 10 generators \cite{hall}. Considering the next oscillator in the chain, we have 3 linearly independent generators from the control of the previous modes, 3 additional ones from the added mode and 4 generators from their commutator (\ref{commutator}). Hence, we have all the necessary generators for complete control of the additional oscillator and its neighbor. The full controllability of the total system now follows by induction in the number of oscillators.


The above arguments prove the controllability of the system but do not specify how to create desired transformations.
Our aim is to drive the chain of oscillators from their ground state to a state
specified by a symplectic transformation matrix $T$. We now describe a simple algorithm to obtain any symplectic matrix $T$ by a sequence of couplings that amounts to factorization 
$T=\prod_i D_{n_i}(S_{n_i})$.

Our strategy is to express the target matrix as $T=\prod_{l=1}^{\tilde{N}}U_l$. The rightmost matrix $U_{\tilde{N}}$ in this product decorrelates the last cradle mode from $T$. We construct $U_{\tilde{N}}$ such that its two last rows are identical to those of $T$. Since these rows determine the effect of $T$ on the last cradle mode, the product $TU_{\tilde{N}}^{-1}$ will have the last mode decorrelated%
\footnote{Analogously, one may construct a matrix $V_{\tilde{N}}$ that has the last two columns identical to $T$. 
Then the product $V_{\tilde{N}}^{-1}T$ will decorrelate the last mode.}. The block-diagonal form of matrices $D_n$ makes the construction of $U_{\tilde{N}}$ simple, 
as each $2\times 2$ block can be created by a product of the form $D_n(S_1)D_{n-1}(S_2)D_n(S_3)$. This product contains all linearly independent generators coupling modes $n$ and $n-1$, thus guaranteeing the existence of a solution to the appearing equations. The equations can then easily be solved by computer algebra to determine the corresponding sequence $\{ \Omega_{n_i}(t)\}$ of couplings. After $\tilde{N}$ such three-step sequences, we find the required $U_{\tilde{N}}$ and thus a new target matrix $T'=TU_{\tilde{N}}^{-1}$ which has its dimensions lowered by 2. We repeat this procedure for the subsequent modes, until all the modes are decorrelated, each being in the ground state. With this algorithm, any symplectic matrix $T$ can be produced in at most  $3N(N-1)/2$ steps.

This procedure is reminiscent of a triangular arrangement of $N(N-1)/2$ two-mode beam splitters to create any desired $N$-mode beam splitter~\cite{reckzeilinger}. In conjunction with single-mode squeezers, such an arrangement allows the construction of any symplectic transform via a physical realization of the Bloch-Messiah decomposition~\cite{blochmessiah}. In contrast, our procedure is more suited for a harmonic chain and does not give an equivalent matrix decomposition. Non-classical correlations are created in every step of the transformation, as we illustrate below. 

\begin{figure}
\includegraphics[width=1\columnwidth]{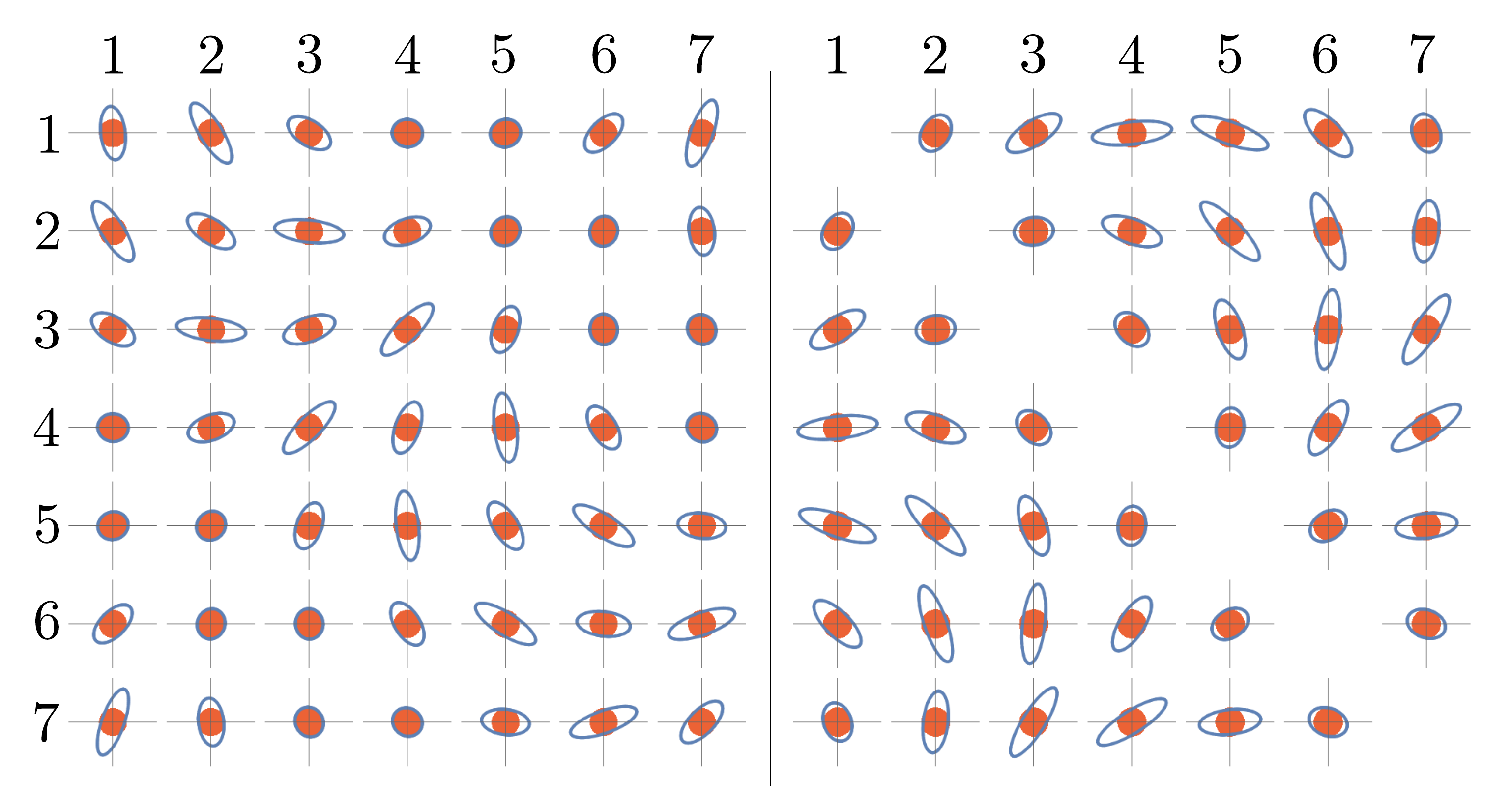}
\caption{Quadrature uncertainties in phase space $(x,p)$ for all pairs $n,m$ of oscillators in a chain of length $N=7$:  Left panel shows the uncertainties of the sum coordinates $\frac{1}{\sqrt{2}}\left(\hat{x}_n+\hat{x}_m,\hat{p}_n+\hat{p}_m\right)$ with the single-oscillator uncertainties on the diagonal; Right panel shows the uncertainties of the difference coordinates $\frac{1}{\sqrt{2}}\left(\hat{x}_n-\hat{x}_m,\hat{p}_n-\hat{p}_m\right)$. 
Solid (orange) circles are the uncertainty of the mode in its vacuum state for scale. Ellipses (blue) are the uncertainties for the squeezed $k=1$ pseudo-phonon state with $\xi=1$, see Eq. (\ref{phononstate}). }
\label{phasespace}
\end{figure}

To demonstrate the possibilities of Gaussian state engineering, we now use a specific example. 
Consider the pseudo-phonon modes in a finite chain of coupled harmonic oscillators,
\begin{linenomath*}
\begin{align}
\tilde{a}_k=\frac{1}{\sqrt{N}}\sum_{n=1}^Ne^{i\frac{2\pi k}{N} n}\hat{a}_n,
\end{align}
\end{linenomath*}
where $\hat{a}_n=\frac{1}{\sqrt{2}}(\hat{x}_n+i\hat{p}_n)$ are the single-site annihilation operators and $k \in \{-N/2+1,\ldots,N/2\}$ for even $N$, or $k \in \{-(N-1)/2,\ldots,(N-1)/2\}$ for odd $N$. 
The excitation energy of mode $\tilde{a}_k$ is uniformly distributed across the entire chain, while the phase difference between any two neighboring oscillators is given by the ``crystal momentum'' $2\pi k/N$. The mode $\tilde{a}_{k=0}$ corresponds to the total displacement which is inaccessible to our manipulations.
 
Let us realize the transformation
\begin{eqnarray}\label{phononstate}
\tilde{a}_{k_{1,2}} &=& \cosh(\xi)\tilde{a}_{k_{1,2}}^{(0)}-i\sinh(\xi)(\tilde{a}_{k_{2,1}}^{(0)})^\dag,\\
\tilde{a}_{k} &=& \tilde{a}_k^{(0)},\quad\mathrm{for}\quad k\neq k_1,k_2 , \nonumber
\end{eqnarray}
where the superscript $(0)$ denotes the Heisenberg operators for the uncoupled system in the ground state. 
Transformation (\ref{phononstate}) corresponds to the unitary evolution of the initial vacuum state under the effective Hamiltonian 
$H_\mathrm{\mathrm{eff}}=J(\tilde{a}_{k_1}\tilde{a}_{k_2}+\tilde{a}_{k_1}^\dag\tilde{a}_{k_2}^\dag)$
for time $\xi/J$. This Hamiltonian describes resonant production of correlated phonon pairs, similar to creation of photon pairs in optical parametric amplification~\cite{pierre}. Analogous processes occur in collisions of ultra-cold atoms~\cite{kheruntsyan}, photon-phonon entanglement in optomechanical systems~\cite{optomechent}, dynamics of driven Bose-Einstein condensates~\cite{drivenBEC} and the dynamical Casimir effect~\cite{casimir}.

Equations (\ref{phononstate}) describe an entangled state of two pseudo-phonon modes $k_1$ and $k_2$. For $k_1=k_2$, a single phonon mode is squeezed. Such a non-classical, highly de-localized state is well suited to benchmark our procedure: every oscillator is entangled with every other oscillator in the chain, resulting in $N(N-1)/2$ entangled pairs. In addition, every single oscillator is in a squeezed state. The magnitude of the single-mode squeezing and pairwise entanglement is small, because the entanglement is equally shared among all oscillators. 
In Fig.~\ref{phasespace} we show the uncertainties of the quadrature components for the sum and difference coordinates at different sites of a chain of length $N=7$, for the case of $\xi=1$. 
The quadrature uncertainties of single oscillators are on the diagonal in the left panel of Fig.~\ref{phasespace}. Every individual oscillator is in a squeezed state. The orientation of the squeezing ellipse in phase space, identified with the squeezing phase, is rotated by the crystal momentum $2\pi k /N$ for any two neighboring sites, resulting in the squeezing ellipse rotating $k$ times across the entire system. The sum and difference coordinates of any two oscillators $n$ and $m$ are squeezed, with the squeezing phase and magnitude modulated by the crystal momentum. If the sum coordinate of two oscillators is squeezed weakly, their difference coordinate exhibits strong squeezing (compare the two panels of Fig.~\ref{phasespace}).

\begin{figure}
\includegraphics[width=0.8\columnwidth]{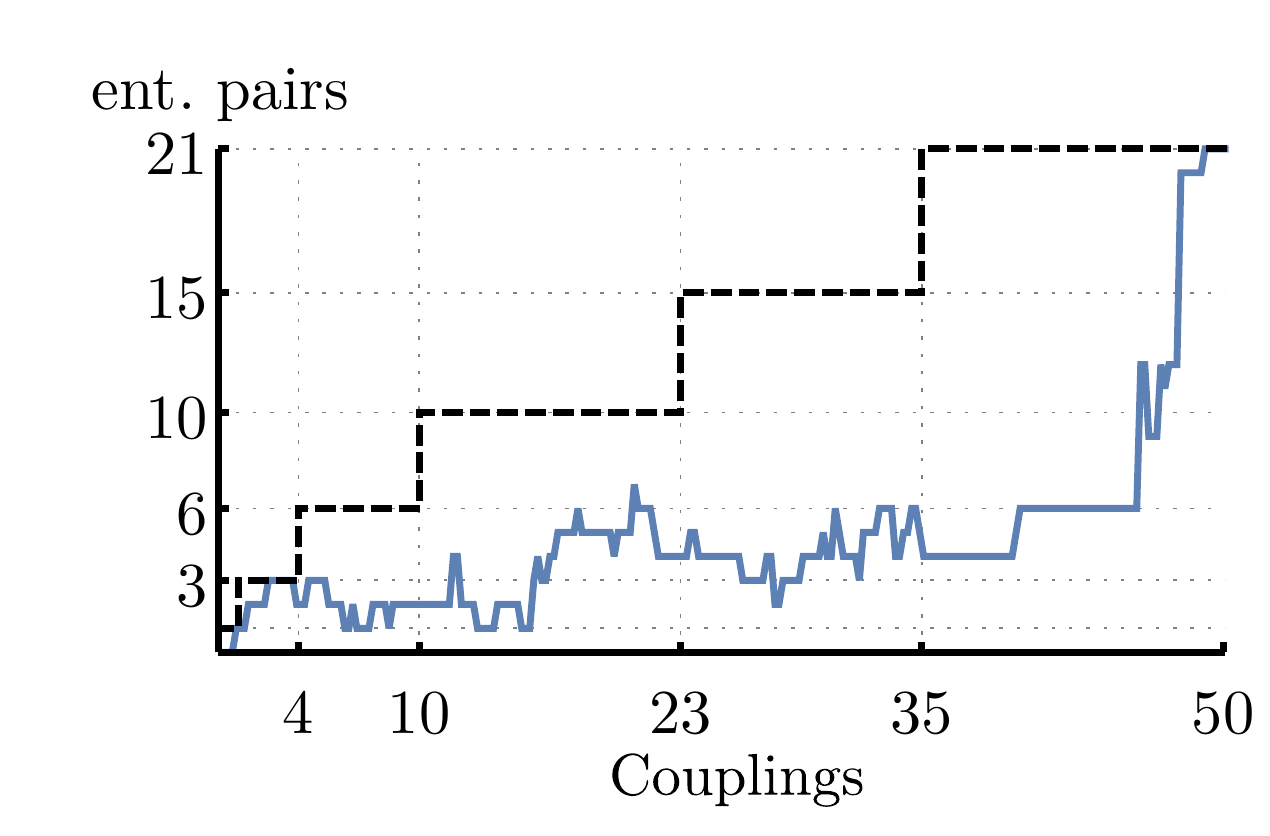}
\caption{Dynamics of preparation of the squeezed phonon state of Fig.~\ref{phasespace}. 
Solid line shows the number of entangled pairs of harmonic oscillators versus the number of applied couplings.
Dashed line gives the maximal number of entangled pairs for a given number of involved oscillators. 
The grid lines on the abscissa give the times of decorrelations of the cradle modes and the grid on the ordinate gives  
the number of entangled pairs $\frac{1}{2} N(N-1)$ for $N$ from 1 to 7.}
\label{dynamics}
\end{figure}

Our algorithm requires 51 sequential couplings $\Omega_{n_i}$ to create the squeezed phonon state for $N=7$, see Fig.~\ref{dynamics}. The procedure increases step by step the number of involved cradle modes, and every subsequent mode requires larger number of couplings for the creation of the desired correlations. Non-classical correlations do not grow monotonically, but occur predominantly during the final stages of the sequence. We note that the sequence found by our algorithm is not unique and relaxing the requirement of non-overlapping couplings will result in more efficient preparation procedures of the desired many-body states. The solution found by our algorithm may then serve as a seed for methods from optimal control theory, subject to a given set of constraints and implementation-specific dissipation mechanisms~\cite{optcontrol}.

A chain of harmonic oscillators can be realized by an array of coupled optomechanical resonators \cite{omarray1} or an ensemble of ultra-cold atoms in an array of microtraps \cite{saffmann2010, browaeys2016,pohl2014}. Controlled couplings between the oscillators can be provided by selective Rydberg dressing of neighboring atoms using non-resonant, amplitude-modulated laser fields~\cite{buchmann2017, rost2010}.
The read-out of the state requires a position measurement of each oscillator. If the oscillator chain is implemented as an optomechanical array, measurement schemes from entangled frequency combs~\cite{freqcombs} can be employed. For oscillator chains consisting of ultracold atoms in a quantum gas microscope, the quadratures can be inferred by freezing the motional state of the system after state-preparation and free evolution or via time-of-flight measurements~\cite{greiner2015, zwierlein2015,kuhr2015, takahashi2016}. Correlation matrices of the shot-to-shot fluctuations can be directly compared to the desired correlations, such as the ones pictured in Fig.~\ref{phasespace}, to verify the engineered state of motion. 

To summarize, we have shown that temporal control of the beam-splitter and parametric couplings between neighboring sites of a chain of $N$ harmonic oscillators gives complete symplectic control over $N-1$ oscillator modes. We have introduced the cradle modes of the system which allowed us to develop an algorithm to produce any desired state using at most $3N(N-1)/2$ couplings between neighboring oscillators. 
We have demonstrated our algorithm by engineering a highly correlated non-local state that appears in a variety of physical systems. 
Our method may also prove useful in the treatment of lattice models for quantum simulators or ensembles of qubits for quantum computation. 
We note before closing that symplectic controllability cannot change the purity of the system, i.e., its effective temperature. Finite temperature physics can be simulated by treating half the modes as an effective reservoir for the remaining system. Thus symplectic controllability of $N-1$ modes translates into complete controllability in the space of mixed Gaussian states of $N/2-1$ modes.

\begin{acknowledgments}
This work was supported by the H2020 FET Proactive project RySQ and by the Villum Foundation.
\end{acknowledgments}


\begin{thebibliography}{}

\bibitem{schrodinger1914} E. Schr\"odinger, Annalen der Physik (Leipzig) {\bf 349}, 916 (1914). 

\bibitem{dudnikova2017} T. V. Dudnikova, J. Math. Phys. {\bf 58}, 043301 (2017).

\bibitem{jeugt2008} S. Lievens, N. I. Stoilova, and J. Van Der Leugt, J. Math. Phys. {\bf 49}, 073502 (2008). 

\bibitem{mingo2003} N. Mingo and Liu Yang, Phys. Rev. B {\bf 68}, 245406 (2003).

\bibitem{dissipative} S. Ma, M. J. Woolley, I. R. Petersen and N. Yamamoto, J. Phys. A: Math. Theor. {\bf 50}, 135301 (2017). 

\bibitem{ionchain}  M. Ramm, T. Pruttivarasin and H. H\"affner, New J. Phys. {\bf 16}, 063062 (2014).

\bibitem{ionchain2} S. Debnath, N. M. Linke, S.-T. Wang, C. Figgatt, K. A. Landsman, L.-M. Duan, C. Monroe,  arXiv:1711.00216 (2017). 

  
\bibitem{omarray1}  H. Okamoto, A. Gourgout, C. Y. Chang, K. Onomitsu, I. Mahboob, E. Y. Chang, and H. Yamaguchi, Nature Phys. {\bf 9}, 480 (2013). 

\bibitem{waveguide1} R.~J.~Chapman, M.~Santandrea, Z.~Huang, G.~Corrielli, A.~Crespi, M.~-H.~Yung, R.~Osellame and A.~Peruzzo, Nature Commun. {\bf 7}, 11339 (2016). 

\bibitem{omarray2} N.~Spethmann, J.~Kohler, S.~Schreppler, L.~Buchmann and D.~M.~Stamper-Kurn, Nature Phys. {\bf 12}, 27 (2016).

\bibitem{thermodynchains} C.~B.~Mendl, J.~Lu, and J.~Lukkarinen, Phys.~Rev.~E {\bf 94}, 062104 (2016).

  
\bibitem{plenio2002} K. Audenaert, J. Eisert, M. B. Plenio, and R. F. Werner, Phys. Rev. A {\bf 66}, 042327

\bibitem{reznik2004} A. Botero and B. Reznik, Phys. Rev. A {\bf 70}, 052329

\bibitem{brukner2006} J. Kofler, V. Vedral, M. S. Kim, and C. Brukner, Phys. Rev. A 73, 052107 (2006).

\bibitem{plenio2004} M.~B.~Plenio, J.~Hartley and J.~Eisert, New J.~Phys.~{\bf 6}, 36 (2004).

\bibitem{tonni2017} A.~Coser, C.~de Nobili and E.~Tonni, J. Phys. A {\bf 50}, 314001 (2017).

\bibitem{vitali2015} S. Zippilli, J. Li, and D. Vitali, Phys. Rev. A {\bf 92}, 032319 (2015). 

\bibitem{morigi2014} B. G. Taketani, T. Fogarty, E. Kajari, Th. Busch, and G. Morigi, Phys. Rev. A {\bf 90}, 012312 (2014). 

\bibitem{giorgos} G. Nikolopoulos and I. Jex, {\it Quantum state transfer and Network Engineering}, Springer Berlin, Heidelberg (2014).

\bibitem{plenio2006} M.~J.~Hartmann, M.~E.~Reuter and M.~B.~Plenio, New J. Phys. {\bf 8}, 94 (2006).

\bibitem{plenio2005} M. B. Plenio and F. L. Semiao, New J. Phys. {\bf 7}, 73 (2005). 

\bibitem{plenio2005_2} A. Perales and M. B. Plenio, J. Opt. B: Quantum Semiclass.~Opt. 7 S601 (2005).

\bibitem{gaussqis} C. Weedbrook, S. Pirandola, R. Garcia-Patron, N. J. Cerf, T. C. Ralph, J. H. Shapiro, and S. Lloyd, Rev. Mod. Phys. {\bf 84}, 621 (2012). 

\bibitem{cluster} C. C. Tison, J. Schneeloch, P. M. Alsing, arXiv:1708.07588.  


\bibitem{universalCVqcomp} N.~C.~Menicucci, P.~van Loock, M.~Gu, C.~Weedbrook, T.~C.~Ralph, and M.~A.~Nielsen, Phys.~Rev.~Lett.~{\bf 97}, 110501 (2006). 

\bibitem{freqcombs} Y. Cai, J. Roslund, G. Ferrini, F. Arzani, X. Xu, C. Fabre and N. Treps, Nature Communications {\bf 8}, 15645 (2017).

\bibitem{cloning} Fr\'ed\'eric Grosshans and Philippe Grangier, Phys. Rev. A {\bf 64}, 010301(R) (2001). 

  
\bibitem{saffmann2010} C.~Knoernschild, X.~L.~Zhang, L.~Isenhower, A.~T.~Gill, F.~P.~Lu, M.~Saffman, and J.~Kim, Appl. Phys. Lett. {\bf 97}, 134101 (2010).

\bibitem{browaeys2016} D. Barredo, S. de Leseleuc, V. Lienhard, T. Lahaye, A. Browaeys, Science 10.1126/science.aah3778 (2016).

\bibitem{pohl2014} T. Macri and T. Pohl, Phys. Rev. A {\bf 89}, 011402(R) (2014). 

\bibitem{buchmann2017} L. F. Buchmann, K. M\o lmer, and D. Petrosyan, Phys. Rev. A {\bf 95}, 013403 (2017).

  
\bibitem{martin2012} J. D. Carter, O. Cherry, and J. D. D. Martin, Phys. Rev. A {\bf 86}, 053401 (2012).

\bibitem{spreeuw2017} J. Naber, S. Machluf, L. Torralbo-Campo, M. L. Soudijn, N. J. van Druten, H. B. van Linden van den Heuvell, and R. J. C. Spreeuw, J. Phys. B  {\bf 49}, 094005 (2017). 

\bibitem{genoni1} M.~G.~Genoni, A.~Serafini, M.~S.~Kim, and D.~Burgarth, Phys. Rev. Lett. {\bf 108}, 150501 (2012).

\bibitem{genoni2} U.~Shackerley-Bennett, A.~Pitchford, M.~G.~Genoni, A.~Serafini and D.~K.~Burgarth, Journal of Physics A: Mathematical and Theoretical, {\bf 50}, 155203 (2017). 
  
\bibitem{hall} ``Lie groups, Lie algebras, and representations'', B.~C.~Hall, Springer International (2003).

 \bibitem{optcontrol} J. Werschnik and E. K. U. Gross, J. Phys. B{\bf 40}, R175 (2007). 

\bibitem{reckzeilinger} M. Reck, A. Zeilinger, H. J. Bernstein, and P. Bertani, Phys. Rev. Lett. {\bf73}, 58 (1994). 

\bibitem{blochmessiah} S.L. Braunstein, Phys. Rev. A {\bf 71}, 055801 (2005). 

\bibitem{pierre} ``Elements of Quantum Optics'', P.~Meystre and M.~Sargent, Springer Berlin, Heidelberg(1990). 

\bibitem{kheruntsyan} V. Krachmalnicoff {\it et al.}, Phys. Rev. Lett. {\bf 104}, 150402 (2010).

\bibitem{optomechent} R. Riediger {\it et al.}, Nature {\bf 530}, 313 (2016).

\bibitem{drivenBEC} S. Robertson, F. Michel, and R. Parentani, Phys. Rev. D {\bf 95}, 065020 (2017).


\bibitem{casimir} I. Carusotto, R. Balbinot, A. Fabbri and A. Recati, Eur. Phys. J. D {\bf 56}, 391 (2010). 

  
\bibitem{rost2010} S. W\"uster, C. Ates, A. Eisfeld, and J. M. Rost, Phys. Rev. Lett. {\bf 105}, 053004 (2010).
  
\bibitem{greiner2015} M. F. Parsons, F. Huber, A. Mazurenko, C. S. Chiu, W. Setiawan, K. Wooley-Brown, S. Blatt, and M. Greiner, Phys. Rev. Lett. {\bf 114}, 213002 (2015). 

\bibitem{zwierlein2015} L. W. Cheuk, M. A. Nichols, M. Okan, T. Gersdorf, V. V. Ramasesh, W. S. Bakr, T. Lompe, and M. W. Zwierlein, Phys. Rev. Lett. {\bf 114}, 193001 (2015).

\bibitem{kuhr2015}  E. Haller, J. Hudson, A. Kelly, D. A. Cotta, B. Peaudecerf, G. D. Bruce, and S. Kuhr, Nature Phys. {\bf 11}, 738 (2015). 

\bibitem{takahashi2016}  R. Yamamoto, J. Kobayashi, T. Kuno, K. Kato, and Y. Takahashi, New. J. Phys. {\bf 18}, 023016 (2016). 


\end{thebibliography}
\end{document}